# Magnetic field-dependent resistance crossover and anomalous magnetoresistance in topological insulator $Bi_2Te_3$

Anand Nivedan,[1] Kamal Das,[2] Sandeep Kumar,[1] Arvind Singh,[1] Sougata Mardanya,[2] Amit Agarwal,[2,†] and Sunil Kumar[1,*]

[1]*Department of Physics, Indian Institute of Technology Delhi, New Delhi - 110016, India*
[2]*Department of Physics, Indian Institute of Technology Kanpur, Kanpur - 208016, India*
*Email: †amitag@iitk.ac.in; \*kumarsunil@physics.iitd.ac.in*

We report a metal-insulator like transition in single-crystalline 3D topological insulator $Bi_2Te_3$ at a temperature of 230 K in the presence of an external magnetic field applied normal to the surface. This transition becomes more prominent at larger magnetic field strength with the residual resistance value increasing linearly with the magnetic field. At low temperature, the magnetic field dependence of the magnetoresistance shows a transition from logarithmic to linear behavior and the onset magnetic field value for this transition decreases with increasing temperature. The logarithmic magnetoresistance indicates the weak anti-localization of the surface Dirac electrons while the high temperature behavior originates from the bulk carriers due to intrinsic impurities. At even higher temperatures beyond ~230 K, a completely classical Lorentz model type quadratic behavior of the magnetoresistance is observed. We also show that the experimentally observed anomalies at ~230 K in the magneto-transport properties do not originate from any stacking fault in $Bi_2Te_3$.

## I. INTRODUCTION

Charge transport on the surface of three-dimensional topological insulators (3D-TIs) has attracted great attention in recent years [1-5]. They host insulating bulk bands along with two-dimensional semi-metallic surface states having spin-momentum locked Dirac Fermions [6-9]. Hence, one expects characteristic differences in the electrical and optical responses between systems having Fermi level in the bulk band as opposed to systems having Fermi level in the bulk gap region. From an experimental perspective, a major difficulty is in separating the contribution of the surface states from that of the overwhelming bulk states. This difficulty is further compounded by the presence of unavoidable vacancies and/or anti-site defects, which leads to free carrier concentration and the sample thickness-dependence of the observed physical properties [10]. Even in pristine samples grown under the same growth conditions, significant and different amount of impurities are always present. Often, for 3D-TIs with a narrow band-gap in the bulk such as bismuth telluride ($Bi_2Te_3$) (or bismuth selenide), as per the Mott criterion [11,12], the Fermi level is displaced from the bulk band-gap region into the conduction or the valence band regions due to unintentional bismuth or tellurium (selenium) vacancies, respectively. Therefore, depending on the intrinsic vacancies and anti-site defects level, the as-grown 3D-TIs can show metallic behavior [13-16] as against the expected semiconducting behavior with negligible impurities [17,18]. This poses a challenge in extracting or distinguishing the Dirac's metal-like surface contribution reliably from the overall conductivity behavior [18] in the standard resistivity vs temperature measurements.

One work, around this problem, is to explore the gate-voltage controlled carrier concentration change in samples with a high value of intrinsic vacancies and anti-site defects [19-21]. Another very important aspect from the experimental point of view is to consider the right thickness of the samples for physical property measurements. For samples with large bulk to surface ratio, the bulk carriers dominate. While for ultra-thin 3D-TI films, the surface states from the top and bottom surfaces can hybridize losing their topological invariance. For the latter case, a gap opens up in the otherwise massless Dirac-like surface density of states [22,23]. Tuning of the Fermi level in bulk 3D-TIs can be achieved externally by either chemical doping, or using a back gate, such that it can be moved from the conduction band into the valence band through the bulk gap region and vice-versa [17,21]. In such situations, metallic resistance behavior due to the massless Dirac surface states can be sensitively captured and differentiated from the bulk semiconducting behavior. As a consequence, the temperature dependent resistivity shows a crossover from the metallic behavior at low temperatures to semiconducting behavior beyond a certain temperature [21,24-26]. Since the electronic band structure of 3D-TIs is susceptible to external perturbations, researchers are motivated to find other and better ways to achieve the above goals. A tunable external magnetic field can be a tool in this regard to manipulate the surface states of a 3D-TI, which can give crucial insights about the surface states from the resistance vs temperature behavior itself. The strength of the external magnetic field has to be compared with that of the intrinsic spin-orbit coupling in these materials such that topological invariance loses relevance in a controlled manner.

Lately, the presence of weak anti-localization (WAL) at low magnetic fields in the magnetoresistance of 3D-TIs has also given a new direction in the study of their linearly dispersing semi-metallic surface states. Linear magnetoresistance (LMR) at high magnetic fields and WAL at low magnetic fields have been extensively used for separating out the surface contribution from the bulk [18,22].

For instance, a non-saturating LMR at high magnetic fields was reported in $Bi_2Te_3$ [15,24,25,27-30], $Bi_2Se_3$ [4,14,31,32] and in other topological insulators [26,33]. Additionally, logarithmic magnetoresistance due to WAL [34] at low magnetic fields and low temperatures has also been reported in several TI's [17,20,22,28,35-37] which becomes less dominant at higher temperatures. Now, in many cases, the LMR is accompanied by quantum oscillations, while in the similar number of other cases, no such oscillations were observed [23,26,33]. This is a deviation from the well-known Abrikosov's theory [38,39]. For physically inhomogeneous samples, Little and Parishwood provided an alternate classical theory [40,41] which describes the non-oscillatory linear MR behavior observed in narrow or zero bandgap systems like TIs [14,29,32]. Weak anti-localization in the linearly dispersing surface states in TIs is typically included via the Hikami-Larkin-Nagaoka (HLN) model [34]. However, for bulk samples where the surface carriers are relatively much smaller, HLN model does not yield realistic value of the model fit parameter as expected from the HLN model. Instead, a large deviation is seen, which is due to the presence of multiple bulk conducting channels in the system [17,25,42]. For magnetoconductance at high magnetic fields, HLN model should be appropriately modified to account for the effect of spin-orbit scattering, elastic scattering and other classical phenomena.

respectively, and the bending $E^2_g$ mode at frequency of ~102 cm$^{-1}$. (f) The bulk BZ and its projection on the (001) surface. The relevant high symmetry points are marked. (g) Electronic band structure including SOC along selected high symmetry direction. The Bi-p is denoted by the blue circles, while green circles represent the Te-p character. (h) Band structure of the (001) surface of $Bi_2Te_3$. The shaded blue region marks projected bulk bands, and sharp lines identify the surface states.

In this paper, we present new insights from the temperature and magnetic field dependent resistance properties of a bulk $Bi_2Te_3$ single crystal. Resistance measurements in a large temperature range varying from very low temperature (<10 K) to very high temperature (> room temperature) performed under the application of an external normal magnetic field reveal a competition between the surface and the bulk conducting channels. Although the sample shows metallic behavior at low temperatures, the effect of thermally activated carriers is evident from metallic to insulating-like transition at ~230 K which becomes more prominent with increasing strength of the external magnetic field. Additionally, MR measurements at various sample temperatures provide a complementary view of the underlying physics of the surface states which have been invoked while interpreting the experimental results at low temperatures. At low temperatures and high magnetic fields, a non-saturating LMR is evident while at high temperatures and all the experimental magnetic fields, a quadratic MR behavior is observed. The demarcation temperature which distinguishes LMR from the quadratic MR behavior is again same (~230 K) indicating that the magnetic field induced metal to insulator-like transition in resistance vs temperature plots also has some surface states contribution.

## II. CRYSTAL AND ELECTRONIC STRUCTURE

Bismuth telluride crystallizes in the rhombohedral crystal lattice with space group $R\bar{3}m$ (No.166) [43]. The hexagonal conventional unit cell of $Bi_2Te_3$ is shown in Fig. 1(a). The unit cell consists of six Bi atoms placed at Wyckoff position 6c, six Te atoms at 6c, and three Te atoms at position 3a. These atoms form a stacking of hexagonal layers along the hexagonal c-direction. The stacking sequence can be thought of as a repetition of five atomic layers, typically known as a quintuple layer (QL) [see Fig. 1(b)-(c)]. High quality single crystals of $Bi_2Te_3$ were used for all the experiments reported in this paper. The typical lateral size of the uniform surface of the samples used was approximately 5 mm × 3 mm. X-ray diffraction (XRD) and Raman scattering measurements were used to characterize them, and the corresponding results are shown in Figs. 1(d) and 1(e), respectively. The presence of only (00$l$) peaks in the XRD pattern confirms a c-axis oriented crystal. The narrow peaks seen in the XRD pattern at 2θ angles of ~17.40°, 44.50°, 54.20°, and 64.2° with their comparable peak-widths of <0.4 degrees have been identified in Fig. 1(d) as the characteristic single-crystalline peaks from (006), (0015), (0018) and (0021) planes. [44,45] The Raman active phonon modes corresponding to stretching modes of $A^1_{1g}$ and $A^2_{1g}$

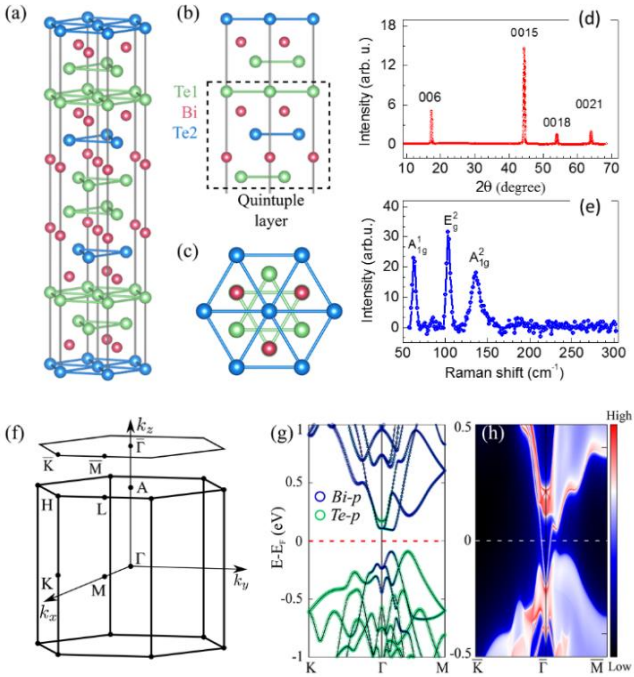

FIG. 1. (a) Hexagonal unit cell of $Bi_2Te_3$ crystal structure. (b) One quintuple layer and (c) the (001) view of the unit cell. (d) XRD intensity peaks at various angles depict highly oriented crystalline structure of bulk $Bi_2Te_3$ sample used in the experiments. (e) Raman spectra showing the stretching $A^1_{1g}$ and $A^2_{1g}$ modes at frequencies of ~62 cm$^{-1}$ and 137 cm$^{-1}$,

appear at ~62 cm$^{-1}$ and ~137 cm$^{-1}$, respectively, and the bending mode of E$^2_g$ appears at ~102 cm$^{-1}$, as seen in Fig. 1(e). These Raman peaks are placed exactly at the expected frequencies, and their line widths are consistent with the reported Raman spectra for single-crystalline pristine structure [45-47].

Because of the dominant covalent character, the intralayer bonding within the QLs is much stronger than the Van-der-Waal type inter-QL interaction. The bulk Brillouin zone (BZ) and the projected BZ of the (001) surface are shown in Fig. 1(f). The electronic band structure calculated along selected directions in the bulk BZ is shown in Fig. 1(g). The orbital-weighted band structure confirms the presence of a band inversion around the high symmetry Γ-point. To check the topological character of this system, we construct a semi-infinite slab of Bi$_2$Te$_3$ along the (001) direction and calculate its surface state spectrum. The calculated surface state spectrum is shown in Fig. 1(h), where the presence of non-trivial linearly dispersing massless Dirac like surface states, within the bulk gap, is evident.

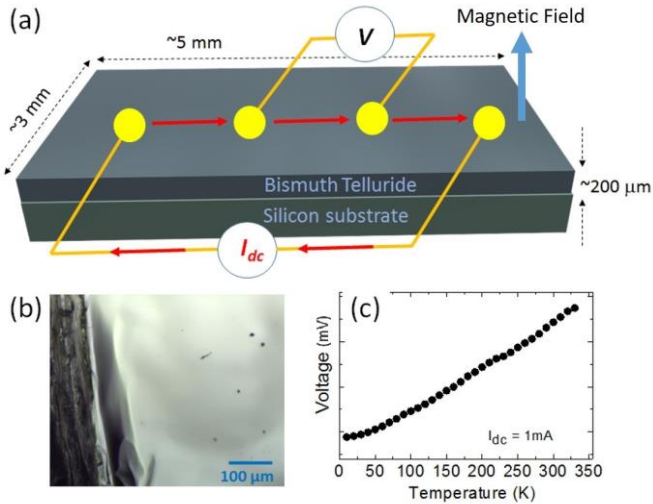

FIG. 2. (a) Schematic of the experimental arrangement for four-probe electrical measurements during R-T and MR experiments. The sample dimensions are also mentioned. (b) Optical image of the single-crystalline bulk Bi$_2$Te$_3$ sample used in all the experiments reported in this paper. (c) A representative result from the temperature-dependent resistance measurement using four-probe geometry at a fixed dc bias current.

## III. EXPERIMENTAL DETAILS AND TRANSPORT MEASUREMENTS

For the temperature-dependent resistance (R-T) and magnetoresistance (MR) measurements reported in this paper, we have used a freshly peeled sample of dimensions ~5 mm x 3 mm x 0.2 mm and placed it inside a physical property measurement system (PPMS, Quantum Design). A schematic of the four-probe measurement geometry is shown in Fig. 2(a) while Fig. 2(b) shows an optical image of the clean and nearly uniform surface of the sample on which the physical property measurements were performed. The left edge of the sample is also seen in the optical image, indicative of the stacking of the layered structure from which a thin leaf can be easily peeled-off for fresh measurements. All the R-T measurements were carried out in the four-probe geometry with equally spaced contacts on the clean surface of the freshly cleaved sample. For a fixed dc bias current of ~1mA between the two outer contacts, voltage across the two inner contacts was monitored while varying sample temperature. Figure 2(c) shows a representative result of the temperature-dependent resistance measurement without any external magnetic field taken in the temperature range of 10-330 K. The R-T measurements were also carried out under the application of a constant external magnetic field normal to the surface.

Similarly, the magnetic field dependent resistance measurements at a few specific temperatures, i.e., the R-B isotherms were taken for magnetic fields applied perpendicular to the surface and varying in the range from -7 Tesla to +7 Tesla. The magnetoresistance was calculated using the relation, $\Delta R/R(0) = [R(B) - R(0)]/R(0)$ where $R(B)$ and $R(0)$ are the resistances measured in the presence and absence of the magnetic field, respectively.

## IV. TEMPERATURE DEPENDENCE OF RESISTANCE AT VARYING MAGNETIC FIELDS

We show the experimentally measured temperature dependent resistance in Fig. 3(a) for three different magnetic field values, applied perpendicular to the measurement surface. The continuous lines in Fig. 3(a) are theoretical fits whose details are discussed later in this section. Clearly, the low-temperature behavior for all four magnetic field values is metallic, while the high-temperature behavior shows a deviation from the metallic nature beyond a certain temperature. The *R-T* curve in Fig. 3(a) for magnetic field 6.5 T shows a clear transition from metallic to non-metallic behaviour at temperature of ~230 K. Without any external magnetic field but by changing the Fermi level either by chemical doping or external bias, a similar temperature dependent transition has been reported in various other 3D-TIs such as Bi$_2$Se$_3$ [19,21], and Bi$_2$Se$_2$Te [26] including Bi$_2$Te$_3$ [25,29,30].

In addition to the appearance of oscillations on the R-T curves with the increasing magnetic field, the residual resistance R$_r$ value near the zero temperature also increases upon increasing the magnetic field as shown in the inset of Fig. 3(a). Apparently, experiments that reveal both of these features, i.e., the appearance of magnetic field-dependent oscillations in the R-T data as well as the residual resistance R$_r$ in Fig. 3(a) have not been reported hitherto in any of the 3D-TI systems. These clearly indicate a magnetic field induced effect to be the origin of these features which need detailed theoretical investigations. To visualize the oscillatory features and the metal to insulating-like transition in the R-T data at ~230 K, we plot the differential of the same in Fig. 3(b) (note that we have shifted the consecutive data sets vertically for better visualization). The transition temperature of ~230 K marked by the vertical dashed line in Fig. 3(b), irrespective of the amount of the magnetic field, is more distinguishable now.

From Fig. 3(c), where $\Delta R(T)|_B = R(T)|_B - R(T)|_{B=0}$ has been plotted, the effect of the magnetic field B on the metal-insulator like transition can be seen with better clarity. Ignoring the small oscillations in the data, $\Delta R(T)|_B$ remains nearly constant below the transition temperature of ~230 K. In this regime, the role of the increasing magnetic field is seen only to increase the residual resistance, $R_r$. We also see from Fig. 3(c) that above the transition temperature, $\Delta R(T)|_B$ falls off rapidly with the temperature and the fall is much sharper at larger magnetic fields. It is also evident from Fig. 3(c) that all of the $\Delta R(T)|_B$ curves are approaching near zero-value at a common temperature between 350 K and 400 K, independent of the applied magnetic field value.

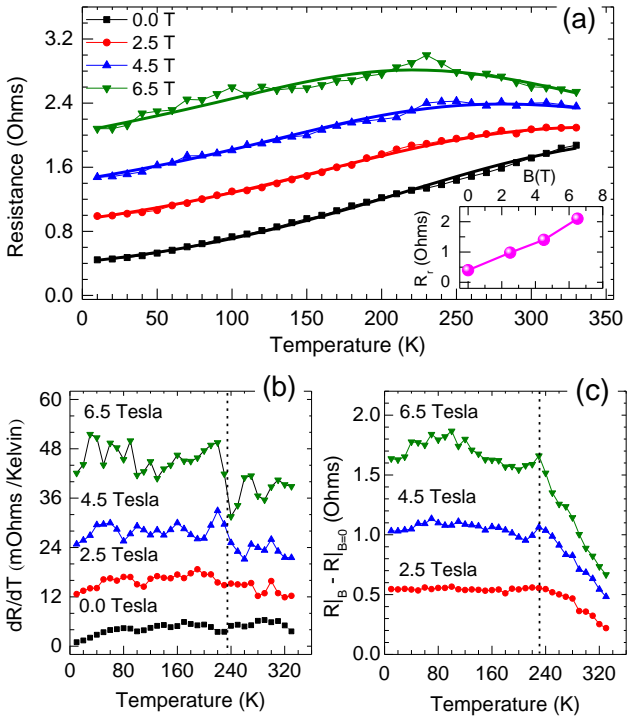

FIG. 3. (a) Temperature dependent resistance plots for $Bi_2Te_3$ at various magnetic field values. The inset shows the variation of the residual resistance $R_r$ vs B(T). (b) Differential plots of same R-T graph at 0.0, 2.5, 4.5 and 6.5 Tesla which have been shifted vertically by constant values for clarity. (c) $\Delta R(T)|_B = R(T)|_B - R(T)|_{B=0}$, plots of the same R-T graphs at 2, 4.5 and 6.5 Tesla. $R(T)|_{B=0}$ is the resistance without any magnetic field.

The insulating bulk bandgap in $Bi_2Te_3$ is of ~0.2 eV (~2400 K). The metallic nature in the R-T curves is due to the contributions from the intrinsic impurity/vacancy concentration in the bulk and the conducting surface states while the insulating behavior is solely due to the bulk states. The competition between the impurities and surface state driven metallic character and the bulk insulating behaviors, and their intricate dependence on the applied magnetic field, gives rise to the observed temperature dependence of the R-T data.

The observed anomaly in the magnetic field induced correction in the temperature dependent resistance, $\Delta R(T)|_B$ and the peculiar behavior at ~230 K in the R-T plot of Fig. 3(a) are quite intriguing and need a quantitative understanding. Around similar temperature value, light-induced photocurrent has been seen to switch sign from negative to positive value [2]. Few other studies in the literature on 3D-TIs have also reported an anomaly at ~230 K, for example, in the thermal expansion coefficient and coherent phonon amplitudes in $Sb_2Te_3$ [16,48]. The recent article by Prakash et al., [48] suggested that the origin of this effect can be the presence of stacking faults developed in the sample. However, our *ab-initio* calculations for $Bi_2Te_3$, rule out this possibility of temperature-induced transition to a stacking fault state, with no topological order (see Appendix A for details).

The low temperature metallic behavior seen in Fig. 3(a) has a contribution from the Dirac-like massless surface electrons around the Fermi energy, as shown in Fig. 1(h). At low temperatures, the resistance is dominated by elastic impurity scattering while at higher temperatures, it is dominated by the electron-phonon scattering processes [49]. Thus, the surface and the intrinsic bulk vacancies related metallic contribution to the conductance, $G_m$, can be modelled [26] via the following equation,

$$G_m(T) = \frac{1}{a+bT+cT^2} \quad (1)$$

Here, $a$ represents the residual resistivity due to elastic impurity scattering, and it is proportional to the $R_r$ shown in the inset of Fig. 3(a). The other two parameters, $b$ and $c$ depend on various other scattering mechanisms which are also dependent on the external magnetic field strength. For instance, the linear temperature-dependent term $b$ comes from electron-phonon scattering, which is the major source of quasiparticle decay and backscattering [21,49]. It occurs due to breaking of the periodic symmetry present in the lattice system. Phonons are responsible for breaking the discrete lattice symmetry and contribute to the temperature dependence of the resistivity as a result of electron-phonon scattering. Additionally, there can be other mechanisms also contributing to the quadratic dependence of the resistivity on temperature. However, the small value of the fitting parameter $c$ indicates that the electron-phonon scattering term is the most dominant mechanism.

At sufficiently high temperatures, the contribution from the bulk insulating states due to comparatively much more density of states in the bulk, start to dominate. Keeping in mind the magnitude of the bulk band gap $\Delta$ ~ 2400 K in $Bi_2Te_3$, the thermally excited quasiparticles across the bulk gap contribute at relatively higher temperatures leading to the thermally activated bulk conductivity, $G \sim exp(-\Delta_i/T)$, where $\Delta_i$ is the thermal activation energy. The thermal activation is for the transition of the electrons in the impurity and defect states to the conduction band. This contribution to overall conductivity is given by [19,21],

$$G_i(T) = \frac{1}{de^{\Delta_i/T}} \quad (2)$$

Here, $d$ is a constant related to the finite density of states and the charge carrier mobilities.

Considering parallel conduction of the metallic carriers due to surface and bulk vacancies, and the bulk insulating channels, the total temperature-dependent resistance can be obtained to be,

$$R(T) = \frac{1}{G_m(T) + G_i(T)} \quad (3)$$

We have used Eq. (3) for fitting the data in Fig. 3(a) where the fits have been shown with thick solid curves at different field values and corresponding fit parameter values are tabulated in Table I. It is seen that a reasonably good fit can be obtained with the energy-parameter $\Delta_i$ varying very weakly with the magnetic field. This signifies that the transport in $Bi_2Te_3$ occurs through a combination of the surface and intrinsic bulk metallic, and thermally excited insulating bulk states. We find the mean value of $\Delta_i$ to be ~0.1 eV, i.e., about half of the bulk insulating band gap value obtained in DFT calculations. Broadly, the low-temperature dependence of the magnetoconductivity arises from the surface states and shows a metallic behavior. However, at higher temperatures, the thermally activated carriers in the bulk start to dominate leading to an insulating behavior.

TABLE I. The fitting parameters for the R-T curve in Fig. 3(a), based on Eq. (3).

| $B$ (T) | $a$ | $b/10^{-3}$ | $c/10^{-6}$ | $d$ | $\Delta_i$ |
|---|---|---|---|---|---|
| 0.0 | 0.42 | 1.8 | 11.0 | 0.4 | 1090 |
| 2.5 | 0.95 | 2.5 | 8.3 | 0.4 | 1050 |
| 4.5 | 1.45 | 3.0 | 6.9 | 0.4 | 1020 |
| 6.5 | 2.05 | 3.6 | 4.6 | 0.4 | 980 |

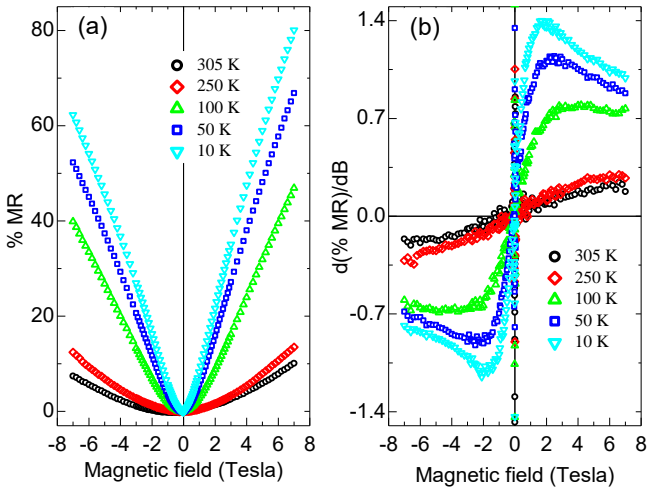

Fig. 4. (a) The measured magnetoresistance (%MR) as a function of the magnetic field B strength. (b) Differential plot of the %MR data of panel (a), for better clarity on the temperature cross-over at a temperature between 100 K and 250 K in the MR behavior.

Overall, the effect of the external magnetic field is two-fold: (i) it induces magnetoresistance through the restricted movement of the surface electrons and the impurity induced bulk carriers and (ii) it breaks the time-reversal symmetry leading to the gapping out of the massless surface states. Note that the values of the parameter '$a$' in Table I at different field strength are same as those of residual resistance $R_r$ given in the inset of Fig. 3(a). This clearly establishes that increasing the magnetic field results in an increase in the scattering cross section of the impurities. Interestingly, we find that the quadratic temperature dependence of $G_m$ (dictated by $c$) decreases with increasing magnetic field strength. In contrast, the coefficient of linear $T$ dependence in $G_m$ (dictated by $b$) increases with increasing magnetic field. These observations from the R-T measurements in the presence of a normal external magnetic field are quite intriguing and open up the possibility of more detailed investigations in future work. One possibility is to explore a Green's function based theoretical approach [50-52].

## V. LINEAR AND QUADRATIC MAGNETORESISTANCE

The experimental results for the magnetoresistance (MR) in percentage $\left(\%MR = \frac{R(B) - R(B=0)}{R(B=0)} \times 100\right)$, taken at a few sample temperatures are shown in Fig. 4(a). In general, we find that for all temperatures, the MR increases with increasing magnetic field strength. Moreover, the overall magnitude of the MR decreases with the increasing temperature. Figure 4(a) shows that there is a clear distinction between the MR data measured at high temperatures (250 K and 305 K) and at low temperatures (10, 50, 100 K). Remarkably, this demarcation temperature is between 100 K and 250 K, similar to the observation of the transition temperature of ~230 K corresponding to the metal to insulator-like transition in the R-T data presented earlier. The MR data at temperatures T > 230 K in Fig. 4(a) shows quadratic magnetic field dependence. However, at low temperatures T < 230 K, it shows a much more complex dependence on the magnetic field. The difference between the behavior of the MR data at T > 230 K and T < 230 K becomes more prominent in Fig. 4(b) where magnetic field differential of the same data, i.e., $d(\%MR)/dB$ are shown. Clearly, the high temperature MR data is quadratic in nature as there is a single slope in the differential data while the same is not true for the data at low temperatures.

In Fig. 5, we show the MR data of Fig. 4(a) separately in the low-temperature regime, i.e., T < 230K in Fig. 5(a) and in the high-temperature regime, i.e., T > 230K in Fig. 5(b), to highlight their distinct behavior. In the low-temperature regime, the data indicates a competition between the logarithmic-type dependence at low magnetic fields and linear dependence at high magnetic fields. These two magnetic field regimes have been separated by the black dashed line drawn across all the magnetic isotherms in Fig. 5 (a). It can be seen that with the increasing temperature, the onset value of the magnetic field for the linear magnetoresistance monotonically decreases. The continuous solid lines in Fig. 5(a) represent the linear fits to the three isotherms at magnetic fields beyond the dashed line. The near logarithmic magnetic field-dependence at low magnetic fields in the low-temperature regime is an attribute of the weak anti-localization due to strong spin-orbit coupling in 2D materials such as 2DEGs. The linearly dispersing Dirac-like conducting states in 3D topological insulators originate from the surfaces states which mimic the 2D nature. Hence WAL behaviour in the MR of 3D-TIs can be attributed to conducting surface states. Similar weak anti-localization behaviour has also been reported earlier in other studies on ultrathin as well as thick films of various 3D-TIs [24,26,28,33,35]. The non-saturating linear (in B) MR at high magnetic fields, is not a unique feature of 3D-TIs. Such a linear MR accompanied by Shubnikov-de Haas oscillations in some cases [23,26,33], has been seen in metals as well where it is typically attributed to open Fermi surface. In contrast, in an ideal 3D-TI, i.e., without any metallicity in the bulk due to the intrinsic impurity (vacancy) induced shifting of the Fermi level, the non-saturating linear MR at high magnetic fields is expected to solely arise due to the conducting surface states. Figure 5(a) demonstrates that the contribution of the surface electrons cannot be ignored completely at the low temperatures even in the presence of comparatively larger bulk metallic contribution.

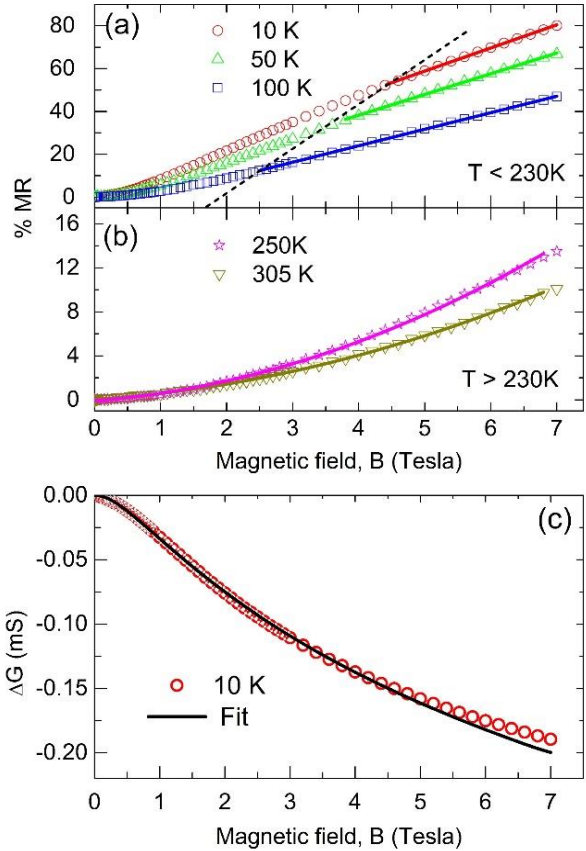

FIG. 5. The MR data (from Fig. 4) plotted here for (a) the low-temperature regime, T < 230 K, and (b) the high temperature-regime, T > 230 K, to highlight different behavior. Continuous lines in (a) are linear and in (b) are quadratic fits. The dashed line in (a) connects the magnetic field values for each isotherm below which the high field dependence deviates from the linear behavior. (c) Magnetic field induced correction in the magnetoconductance, $\Delta G = G(B) - G(B = 0)$, for T = 10 K. The solid black line in (c) shows the fit to the data using HLN model as described in the text.

In the low-temperature regime, there can be two other possibilities for the observation of linear MR at high magnetic field values: (i) MR arising from all the electrons occupying the zero$^{th}$ Landau level as predicted by Abrikosov's quantum theory [23,30,38,39], and (ii) MR arising from the inhomogeneities in the sample as per the classical theory of Parish and Little-Wood [14,29,32,40,41]. We argue that both of these possibilities do not occur in the present case. Here, Abrikosov's theory can be ruled out as we do not observe quantum oscillations indicating the transition of the electron occupancy of the lowest Landau level. Furthermore, the classical origin of LMR owing to inhomogeneities in the sample can also be ruled out as our high-quality crystalline samples were exfoliated from a bulk single crystal, as discussed earlier in Section II. Furthermore, Wang et al., [53] showed that a linear MR could be observed in topological insulators due to surface states even if multiple Landau levels are filled if we include a finite Zeeman splitting. However, it turns out that this kind of MR is quite robust against temperature variations, which is certainly not the case in our experiments.

In the high temperature regime (T > 230 K), a quadratic behaviour of the %MR is seen at all magnetic fields in Fig. 5(b). This can be described well using the classical Lorentz model [35,54,55]. The solid curves in Fig. 5(b) represent the quadratic fit. Again, this feature is not unique to 3D-TIs where bulk states are the main contributors [56-58].

To analyze the experimental observations of low-temperature magnetoresistance in Fig. 5(a), we plot the corresponding magnetoconductance, $\Delta G = G(B) - G(B = 0)$ at the lowest experimental temperature of 10 K, in Fig. 5(c). The weak anti-localization contribution in ΔG due to the Dirac-like surface states in 3D-TIs can be analysed using the following HLN equation [22,24,46],

$$\Delta G = -\alpha \frac{e^2}{2\pi^2 \hbar} \left[ \psi\left(\frac{1}{2} + \frac{\hbar}{4el_\varphi^2 B}\right) - ln\left(\frac{\hbar}{4el_\varphi^2 B}\right) \right] \quad (4)$$

Here, ψ is the digamma function, $l_\varphi$ is the phase coherence length and other constants have their usual meaning. The prefactor α denotes the number of conducting channels, with a value of ½ per conducting channel. Originally, HLN model [34] was given in terms of the magnetoconductivity Δσ for an ideal two-dimensional spin-orbit coupled system (please see Appendix B). The HLN model only accounts for the semi-metallic surface states present in 3D-TIs. Therefore, use of HLN model for 3D-TIs is useful typically for very thin samples [24,28,33,46,59] so that the dominating contribution due to the bulk states can be avoided. Remarkably, this model has also been used in the literature for analyzing magnetoconductance of thick films of 3D-TIs [17,25,42], and it does give a reasonable fit. However, the bulk contribution is typically hidden under the very large value of α compared to its ideal value of ½ per conducting channel. See appendix B for more details on the application of HLN formula to thick samples of 3D-TIs.

Our magnetoconductance data at 10 K fits well with the HLN model for magnetic fields upto 4 Tesla, as shown by the solid black curve in Fig. 5(c). The obtained phase coherence length $l_\phi$ ~ 22.4 nm is broadly consistent with previous experimental results [25,60], however, the value of α ~ $1.3\times10^7$ is very high. Similar high value of α was reported previously for other bulk 3D-TIs [17,25,42] having sample thickness ranging from ~10 μm to ~200 μm. The intrinsic vacancies related bulk metallicity of the sample is the main reason for getting so high value of α. At larger values of the applied magnetic field beyond 4 Tesla, HLN model does not fit the experimental data in Fig. 5(c). A correction in the conductivity due to spin-orbit scattering, phase coherence scattering and classical cyclotronic resistivity terms must be added in the HLN model as the phase coherence length at such high field values no longer remains dominant [33]. Moreover, increasing the temperature of the sample shifts the transport mechanism from the quantum regime to the classical regime. Phase coherence of the surface electrons at high temperatures decreases sufficiently because of inelastic collisions, and therefore the contribution in the overall conductivity is dominated by the bulk transport [33,61].

## VI. CONCLUSION

To summarise, magneto-transport measurements were carried out on layered single-crystalline bismuth telluride, which is a three-dimensional topological insulator. Our R-T measurements in the presence of a perpendicular magnetic field, demonstrate a metal-insulator like transition at a temperature of ~230K. We find a metallic regime at low temperatures followed by an insulating regime at higher temperatures. We modelled this using two-channel conductance mode comprising of (i) metallic surface states and bulk impurities-based carriers and (ii) thermally activated carriers across the insulating bulk gap at higher temperatures.

At low temperatures, the magnetic field dependence of the magnetoconductance shows a transition from logarithmic to linear behaviour where the onset magnetic field value for this transition decreases with the increasing temperature. We showed that the quantum correction to the magnetoconductance at low temperatures is reasonably captured by the weak anti-localization physics within the HLN framework upto 4 Tesla magnetic field with a coherence length of the order of few tens of nm. On the other hand, at high temperatures beyond ~230 K, the dominating bulk states completely mask the traces of the surface states in the magnetoconductance and the classical Lorentz model type quadratic behaviour is observed.

## ACKNOWLEDGEMENTS

SK and AA gratefully acknowledge Science and Engineering Research Board (SERB), and the Department of Science and Technology (DST), Govt. of India for financial support. PPMS facility at IIT Delhi is acknowledged for help during part of the experiments. AN acknowledges Department of Science and Technology, Government of India for INSPIRE fellowship.

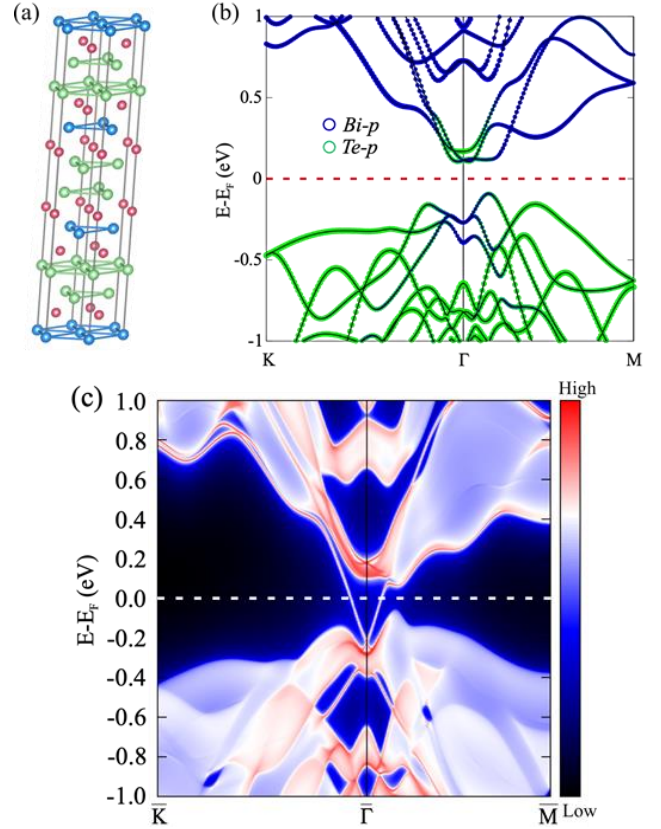

FIG. A.1. (a) Crystal structure of $Bi_2Te_3$ with a stacking fault between the hexagonal layers. (b) Electronic band structure of the faulted structure, where the orbital weight of the Bi-p (blue color) and Te-p (green color) is encoded in the size of the dots. (c) Surface spectrum of faulted $Bi_2Te_3$ along the high symmetry direction of (001) surface Brillouin zone. Clearly, in $Bi_2Te_3$, the topology and the Dirac-like surface states are intact even in the presence of a small tilt.

## APPENDIX A

Here, we show that both the bulk electronic band structure and the surface states in the faulted structure are very similar to that of the pristine structure of $Bi_2Te_3$. In a recent report by Prakash et al. [48] it was shown that the formation of stacking fault in the layered 3D-TI material $Sb_2Te_3$ is crucial to explain an anomaly in its thermal expansion coefficient and few optical properties measured experimentally as a function of the sample temperature. It was shown that $Sb_2Te_3$ undergoes a temperature-induced phase transition from its normal crystalline phase to a topological stacking fault structure which becomes energetically more favourable with increasing temperature. Motivated from this, we have also introduced stacking fault in the basal plane of the hexagonal structure of $Bi_2Te_3$ and explored the electronic and topological properties as summarized in Fig A.1. For ease of calculation, we chose the cell vectors of the faulted structure to be $\mathbf{a}_{SF} = \mathbf{a}_0$, $\mathbf{b}_{SF} = \mathbf{b}_0$, and $\mathbf{c}_{SF} = \mathbf{c}_0 + 1/2\mathbf{a}_0 + 2/3\mathbf{b}_0$, where $\mathbf{a}_0$, $\mathbf{b}_0$, $\mathbf{c}_0$ denote the pristine lattice vectors. The resulting crystal structure is shown in Fig. A.1(a). In contrast to $Sb_2Te_3$, we do not find any notable change in stacking fault induced bulk band structure or in the surface spectrum of faulted $Bi_2Te_3$ [see Fig. A.1(b)-(c)]. Thus, we can rule out the possibility of a stacking fault induced topological phase transition in $Bi_2Te_3$ to be responsible for the observed transport anomaly in our experiments around 230 K.

## APPENDIX B

The logarithmic behavior of the magnetoconductance at low temperatures is an effect of coherent superposition of electrons in the regime where phase coherence length $l_\varphi$ is greater than the mean free path $l_e$ of electrons. At very low temperatures, carrier scattering in spin-orbit coupled systems, including 3D-TIs, leads to destructive quantum interference phenomena. This in turn leads to weak anti-localization of electrons [22,24,36,46] in presence of an external magnetic field, and it manifests as logarithmic corrections in the magnetoresistance. The change in conductivity due to quantum interference in a two-dimensional material system can be described using the HLN model given by Hikami, Larkin and Nagaoka as follows, [34]

$$\Delta\sigma(B) = -\alpha \frac{e^2}{2\pi^2 \hbar}\left[\psi\left(\frac{1}{2} + \frac{\hbar}{4el_\varphi^2 B}\right) - ln\left(\frac{\hbar}{4el_\varphi^2 B}\right)\right]. \quad (B.1)$$

Here, $\Delta\sigma = \sigma(B) - \sigma(B=0)$ is the magnetic field induced quantum correction to magnetoconductivity. The prefactor $\alpha$ in the above equation denotes the number of conducting channels in the system. The absolute value of the parameter α can be either ½ or 1 for single or double coherent channels, respectively. It can be positive or negative depending upon the type of localization. For weak localization, α is negative and for weak anti-localization, α is positive. Weak anti-localization effect is prevalent in those cases where surface contributions are dominant such as in thin samples of 3D-TIs.

As discussed earlier, in real samples, the thickness of the sample dictates the expected value of $\alpha$. For ultrathin samples where thickness is comparable with the coherence length, $\alpha \sim 1$ seems to justify the observations whereas for thicker samples (not bulk), it suggests that $\alpha \sim $ ½ should justify the observations. The HLN model has been routinely used to describe the experimental magnetoconductance of various intrinsic spin-orbit coupled systems at the low temperatures and low magnetic fields. For thicker samples of bismuth telluride 3D-TI, where there are Dirac-like conducting surface states, the problem at hand, is much more involved. In these systems, the direction of the external magnetic field with respect to the crystal axis is an important consideration. Mostly, magnetoconductivity experiments are performed on samples cut along the c-planes, and the external magnetic field applied along the c-direction. In bismuth telluride samples having thickness of more than a few quintuple layers (thickness of each quintuple layer ~1nm) the metallic surface state contribution has been observed [62]. Most of the magnetoconductivity experiments reported in the literature on 3D-TIs have used sample thicknesses much above this limit [24,26-29,46]. In cases where sample thickness is of a few 10's of nm [28,33,46], the HLN model seems to fit the experiments very well with values of $\alpha < 1$ and $l_\varphi$ ~20-200 nm. This is reasonable because in such cases, the bulk contribution in the overall conductivity is limited by the small thickness. In that case, even for samples with bulk metallic nature due to high doping of intrinsic vacancies in 3D-TIs is not an issue as the surface contribution is typically larger. On the other hand, in those cases where the sample thickness is much higher, in microns and beyond, the bulk contribution starts to dominate over the surface electrons [25,42].

Remarkably, it turns out that the 2D HLN formula given in Eq. B.1 also works for thick 3D-TI samples with large bulk contribution as well. Many experimental reports in the literature on thick 3D-TIs [17,25,42], suggest that the magnetoconductivity behaviour at low temperatures and low magnetic fields is different from what can be expected from an entirely classical model. A finite logarithmic trend in the magnetoresistance data in all such cases suggested a distinguishable contribution due to the Dirac's surface electrons over the dominating bulk electrons. In those cases also, the HLN model has been applied with very high values of α (orders of magnitude higher than suggested in the original HLN model) to describe the experimental observations. No additional term in the HLN model was incorporated while fitting the data, and therefore, the high value of α was attributed to the bulk nature of the sample.

A few more things can also be noted from the HLN formula in Eq. (B.1). First of all, the quantity $2e^2/h$ is called quantum conductance, and its SI unit is $\Omega^{-1}$. Therefore, for a dimensionless parameter α having its value varying between -1/2 and 1, $\Delta\sigma$ on the left-hand side in Eq. (B.1) would represent conductance (G) rather than the conductivity (σ). An equivalence between Eq. (4) and Eq. (B.1) can be drawn due to the fact that for atomically thin material systems, the magnetoconductivity and magnetoconductance are not different. Therefore, the apparent inconsistency between the unit $\Omega^{-1}$ of the right-hand side and the unit of conductivity $\Delta\sigma$ on the left-hand side of HLN formula given in Eq. (B.1) is justified for atomically thin or two-dimensional systems. Care must be taken however, when analysing magnetoconductance data from bulk samples using the HLN formula. Use of magnetoconductance should be the right description, however, value of α in that case is justifiable only for extremely thin samples. Otherwise for thicker samples, the parameter α should be accordingly renormalized to account for the finite thickness.

The above confusion might reflect from the large variation of the values of α reported in the literature for 3D-TIs. Though, most of the reports in the literature use Δσ as the magnetoconductivity [20,21,28,35,63], in many other cases, magnetoconductance ΔG is chosen over Δσ while applying the HLN formula on either thin or thick systems [22,24,36,46,64]. Consider the example of typical conductivity of a metallic $Bi_2Te_3$ at room temperature that is in the range of ~$10^2$ $\Omega^{-1}cm^{-1}$ to ~$10^3$ $\Omega^{-1}cm^{-1}$ [15,25,30]. Percentage magnetoresistance (%MR) at field 1 Tesla and temperature of 10 K for a typical $Bi_2Te_3$ or $Bi_2Se_3$ crystals is ~20% [14,15]. Corresponding increment in the conductivity would be around 10% which is too large to be fitted with the HLN model. For instance, with coherence length $l_\varphi$ = 25 nm and $\alpha$ = ½, HLN model gives the conductance of the order of $10^{-6}$ $\Omega^{-1}$. In our paper, we have chosen to use magnetoconductance ΔG (unit $\Omega^{-1}$) for applying the HLN model to fit the data at low temperatures and low magnetic fields. Thus obtained value of $\alpha$ is many orders of magnitude higher than usually reported for extremely thin samples [28,33,46]. It indicates that the bulk contribution in the overall behaviour of the magnetoconductance is many orders higher than the surface contribution even when a small logarithmic nature due to weak anti-localization by surface electrons can be seen in the raw experimental magnetoresistance data (Fig. 4a).